\documentclass[aps,twocolumn,showpacs,superscriptaddress,preprintnumbers]{revtex4}
\usepackage{graphicx}
\usepackage{dcolumn}
\usepackage{bm}

\input{tcilatex}

\begin{document}

\title{In-plane anisotropy of the electronic structure for the
charge/orbital-ordered state in half-doped manganite with layered structure}
\author{Y. S. Lee}
\affiliation{Spin Superstructure Project, ERATO, Japan Science and Technology Agency,
Tsukuba 305-8562, Japan}
\affiliation{Department of Physics, Soongsil University, Seoul 156-743, Korea}
\author{S. Onoda}
\affiliation{Spin Superstructure Project, ERATO, Japan Science and Technology Agency,
Tsukuba 305-8562, Japan}
\author{T. Arima}
\affiliation{Spin Superstructure Project, ERATO, Japan Science and Technology Agency,
Tsukuba 305-8562, Japan}
\affiliation{Institute of Multidisciplinary Research for Advanced Materials, Tohoku
University, Sendai 980-8577, Japan}
\author{Y. Tokunaga}
\affiliation{Spin Superstructure Project, ERATO, Japan Science and Technology Agency,
Tsukuba 305-8562, Japan}
\author{J. P. He}
\affiliation{Spin Superstructure Project, ERATO, Japan Science and Technology Agency,
Tsukuba 305-8562, Japan}
\author{Y. Kaneko}
\affiliation{Spin Superstructure Project, ERATO, Japan Science and Technology Agency,
Tsukuba 305-8562, Japan}
\author{N. Nagaosa}
\affiliation{Correlated Electron Research Center, National Institute of Advanced
Industrial Science and Technology, AIST Tsukuba central 4, Tsukuba 305-8562,
Japan}
\affiliation{Department of Applied Physics, University of Tokyo, Tokyo 113-8656, Japan}
\author{Y. Tokura}
\affiliation{Spin Superstructure Project, ERATO, Japan Science and Technology Agency,
Tsukuba 305-8562, Japan}
\affiliation{Correlated Electron Research Center, National Institute of Advanced
Industrial Science and Technology, AIST Tsukuba central 4, Tsukuba 305-8562,
Japan}
\affiliation{Department of Applied Physics, University of Tokyo, Tokyo 113-8656, Japan}
\date{\today }

\begin{abstract}
We report on the in-plane anisotropy of the electronic response in the
spin/charge/orbital ordered phase of a half-doped layered-structure
manganite. The optical conductivity spectra for a single domain of Eu$_{1/2}$%
Ca$_{3/2}$MnO$_{4}$ unambiguously show the anisotropic charge dynamics which
well corresponds to the theoretical calculation: the optical conductivity
with the polarization along the zigzag ferromagnetic chain direction
exhibits a smaller gap and a larger intensity at lower energies than that of
the perpendicular polarization mostly due to the charge/orbital ordering and
the associated quantum interference effect.
\end{abstract}

\maketitle

Recent studies on strongly correlated electronic systems have revealed many
unexpected phenomena due to interplay among the spin, charge and orbital
degrees of freedom. In particular, nano-scale self-organization and the
associated complexity of electronic systems are issues of current interest.
One of the representative phenomena is the spin/charge/orbital ordering in
transition metal oxides (TMOs), which governs the electrical, optical, and
mechanical properties of the system \cite{Imada98}. Doping carriers into the
parent Mott insulating compound causes a variety of ordering phenomena
including one-dimensional (1D) stripe formation as observed in some layered
Cu- and Ni-oxides. In Mn-oxides near half-doping there appears a
cooperatively ordered pattern of charge, spin, and orbital, referred to as
CE-type magnetic order coupled to the charge/orbital ordering (CO-OO) \cite%
{goodenough55,sternlieb96,murakami98,khomski04,Larochelle05}, which is
involved in the metal-insulator transition with magnetic field and other
various control parameters \cite{orbitalphysics}. Below the CO-OO transition
temperature $T_{CO-OO}$, charges are ordered in a checkerboard-type with
ideally alternating Mn$^{3+}$/Mn$^{4+}$ sites, \textit{i.e.,} the $3d$ $%
e_{g}^{1}$/$e_{g}^{0}$ configuration in addition to localized $t_{2g}$
spins, and $e_{g}$ orbitals are aligned in the zigzag chain type. The
direction of $e_{g}$ orbitals defines the zigzag chain and stripe [Fig.
1(c)]. Below the N\'{e}el temperature $T_{N}$, which is usually far below $%
T_{CO-OO}$, the spins of $3d$ electrons at Mn sites are ordered
ferromagnetically along the orbital zigzag chain path, while
antiferromagnetically between the chains. This well-organized
charge/spin/orbital ordering occurs in the $ab$ plane for a variety of
manganites including layered compounds, and generically has the effective
two-dimensional nature in electronic response even for the pseudo-cubic
perovskite \cite{tobe04}.

To access the attribute of the charge/orbital ordering in more depth, we pay
attention to the in-plane anisotropy of the electronic structure in the
CO-OO phase, in particular, the optical response. The spin order may invoke
the simplest trivial source of the anisotropy. Namely, well below $T_{N}$
where the ferromagnetic chains are coupled antiferromagnetically, the local
electron hopping between the adjacent chains costs a larger energy typically
of the order of $J_{H}S$ ($J_{H}$:Hund's rule coupling, $S$: magnitude of
the $t_{2g}$ spin), than that along the zigzag chain path. Since the optical
processes occur locally, they are insensitive to the range of the magnetic
ordering, and the anisotropy remains even above $T_{N}$. This scenario,
however, is oversimplified and may not apply directly to real experimental
systems that show the drastic change of electronic response below $T_{CO-OO}$%
, which is even well above $T_{N}$. Therefore, we must seriously consider
the effects of CO-OO on the anisotropy of the low-energy optical response.

\begin{figure}[tbp]
\includegraphics[width=0.40\textwidth]{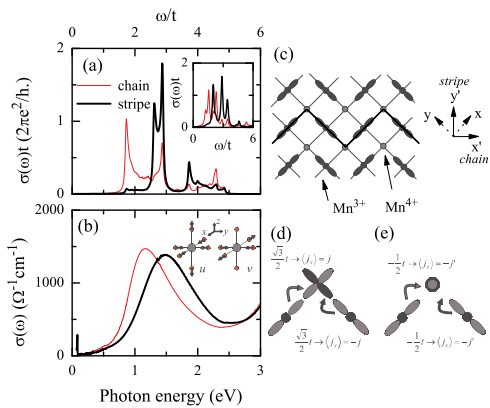}
\caption{(Color online) (a) Optical conductivity spectra along the chain and
the stripe directions from (a) the calculation using the Hartree-Fock method
and (b) the experiment for the single domain of ECMO. In the inset of (a)
the result of the exact diagonalization study is displayed. In the inset of
(b) two Jahn-Teller modes, $u$ and $v$, are drawn with a MnO$_{6}$
octahedron. (c) Schematic diagram for the charge/orbital ordered state at
half-doping. Right bottom panels depict difference in the phases for the
local hopping and current matrix elements from the $d_{3x^{2}-r^{2}}$/$%
d_{3y^{2}-r^{2}}$ orbitals at the electron-rich sites to (d) the $%
d_{x^{2}-y^{2}}$ and (e) the $d_{3z^{2}-r^{2}}$ orbitals at the
electron-poor ones. }
\end{figure}

To elucidate the effects of CO-OO in the in-plane optical response, we take
the following minimal model for a ferromagnetic zigzag chain in the CE
phase; 
\begin{eqnarray}
\mathcal{H} &=&-\sum_{\mathbf{r},\mathbf{r}^{\prime }}\sum_{\alpha ,\beta
=1,2}t_{\mathbf{r},\mathbf{r}^{\prime }}^{\alpha ,\beta }c_{\mathbf{r}\alpha
}^{\dagger }c_{\mathbf{r}^{\prime }\beta }+U\sum_{\mathbf{r}}n_{\mathbf{r}%
1}n_{\mathbf{r}2}  \nonumber \\
&&+E_{JT}\sum_{\mathbf{r}}\left( u_{\mathbf{r}}^{2}+v_{\mathbf{r}}^{2}-2u_{%
\mathbf{r}}(n_{\mathbf{r}1}-n_{\mathbf{r}2})-2v_{\mathbf{r}}(c_{\mathbf{r}%
1}^{\dagger }c_{\mathbf{r}2}+c_{\mathbf{r}2}^{\dagger }c_{\mathbf{r}%
1})\right) .  \label{eq:H}
\end{eqnarray}%
With the annihilation (creation) operators $c_{\mathbf{r}\alpha }$ ($c_{%
\mathbf{r}\alpha }^{\dagger }$) and the number operator $n_{\mathbf{r}\alpha
}=c_{\mathbf{r}\alpha }^{\dagger }c_{\mathbf{r}\alpha }$ of the electron at
site $\mathbf{r}$ with the $e_{g}$ orbital $\alpha $ (=1 or 2 for $%
d_{3z^{2}-r^{2}}$ or $d_{x^{2}-y^{2}}$ respectively), the Hamiltonian given
by Eq.~(\ref{eq:H}) contains the orbital-dependent electron transfer $t_{%
\mathbf{r},\mathbf{r}}^{\alpha ,\beta }$ between the adjacent sites ${%
\mathbf{r}}$ and ${\mathbf{r}^{\prime }}$ with the same spin polarization as
the first term, and the local Coulomb and the Jahn-Teller interactions $U$
and $E_{JT}$ as the second and the third terms, respectively. $u_{\mathbf{r}%
} $ and $v_{\mathbf{r}}$ denote the local Jahn-Teller modes for the MnO$_{6}$
octahedron centered at the Mn site $\mathbf{r}$, as shown in the inset of
Fig.~1~(b). We have neglected the phonon dynamics of these Jahn-Teller
modes. This frozen Jahn-Teller distortion corresponds to the Frank-Condon
approximation which is justified for the strong electron-phonon coupling
case as in the present situation. As for the orbital and directional
dependencies of the transfer integrals, we take the Slater-Koster's value~%
\cite{SlaterKoster} with $V_{dd\sigma }=t=0.25~$eV, which is taken as the
energy unit.

We solve this Hamiltonian within the Hartree-Fock approximation to obtain
the self-consistent values for the expectation values $\langle c_{\mathbf{r}%
\alpha }^{\dagger }c_{\mathbf{r}^{\prime }\alpha ^{\prime }}\rangle $ as
well as the Jahn-Teller distortions $u_{\mathbf{r}}$ and $v_{\mathbf{r}}$.
We note that Hartree-Fock treatment provides a reasonable description of the
ground state even at the large $U$. Since the result is insensitive to the
value of $U$, we may take a typical value for transition-metal ion, $U=8t$.
In the present case of half doping, we have obtained the CO-OO pattern
characterized by the charge disproportionation $\delta n$ and the dominant
orbital character of $d_{3x^{2}-r^{2}}$/$d_{3y^{2}-r^{2}}$ at the
electron-rich ($\frac{1+\delta n}{2})$ sites and $d_{x^{2}-y^{2}}$ at the
electron-poor ($\frac{1-\delta n}{2})$ ones. For this ordering the
Jahn-Teller coupling is indispensable, as previously pointed out \cite%
{horsch04}, and the ground-state properties are insensitive to the choice of 
$U$. Then, we employ the value $E_{JT}=2t$ to reproduce the experimental
result of the optical gap. This choice of $E_{JT}=2t$ gives the charge
disproportionation of $\delta n=0.56$, which is intermediate between the
uniform value $\delta n=0$ for $E_{JT}=0$ and the ideal value $\delta n=1$
for $E_{JT}\rightarrow \infty $.

Using the Hartree-Fock solution, we have calculated the optical conductivity 
$\sigma ^{\mathrm{chain}}$ and $\sigma ^{\mathrm{stripe}}$ along the chain
and the stripe directions, respectively, with the corresponding current
operators $J^{\text{chain}}=(J^{x}+J^{y})/\sqrt{2}$ and $J^{\text{stripe}%
}=(J^{x}-J^{y})\sqrt{2}$ where 
\begin{equation}
J^{x,y}=-\frac{1}{N}\sum_{\mathbf{k},\alpha ,\beta }\frac{\partial t_{%
\mathbf{k}}^{\alpha ,\beta }}{\partial k_{x,y}}c_{\mathbf{k},\alpha
}^{\dagger }c_{\mathbf{k},\beta }
\end{equation}%
with $t_{\mathbf{k}}^{\alpha ,\beta }=\sum_{\mathbf{\delta }}t_{\mathbf{r},%
\mathbf{r}+\mathbf{\delta }}^{\alpha ,\beta }e^{-i\mathbf{k\cdot \delta }}$.
The calculated results for $\sigma ^{\mathrm{chain}}$ and $\sigma ^{\mathrm{%
stripe}}$ in the case of $E_{JT}=2t$ are shown in Fig. 1(a). This
calculation reveals that the optical conductivity along the chain direction
exhibits a smaller optical gap and a larger spectral weight distribution at
lower energies than along the stripe direction. We have also done the
exact-diagonalization study of the same model with the Jahn-Teller
distortion being fixed at the ground state value (Frank-Condon
approximation), and found that the global structure of $\sigma ^{\mathrm{%
chain}}$ and $\sigma ^{\mathrm{stripe}}$ and the emergence of the anisotropy
do not alter even beyond the Hartree-Fock, as shown in the inset of Fig
1(a). The anisotropy originates from the difference in the phase factors for
the combination of $J^{x}$ and $J^{y}$ in the orbital-ordered state and is
an increasing function of the CO-OO order parameters. More explicitly, the
local hopping process from the $d_{3x^{2}-r^{2}}$/$d_{3y^{2}-r^{2}}$
orbitals at the electron-rich sites to the $d_{x^{2}-y^{2}}$ orbital at the
electron-poor site, which is shown in Fig.~1(d), exhibits additive and
negative interference in the chain and the stripe directions, respectively,
leading to the strong anisotropy, while the hopping to the $d_{3z^{2}-r^{2}}$
orbital shown in Fig.~1(e) does not.

To address the in-plane anisotropy experimentally, the measurement for an
in-plane twin-free single domain sample is highly required, but has not been
performed so far due to a great difficulty to obtain a sample large enough
for optical measurement ($\sim $ 100 $\mu $m in case of using microscope
probe). To overcome this technical problem we chose a 2D single layered
structure which is quite suitable for probing the Mn-O plane where the CO-OO
occurs. Due to the two-dimensional electronic nature of the charge/orbital
ordering concerned here, our study with layered manganites is directly
applicable to the case of the pseudo-cubic perovskite. For the well-studied
half-doped material La$_{1/2}$Sr$_{3/2}$MnO$_{4}$ (LSMO), a typical domain
size was found to be $\sim 10$ $\times 10$ $\mu $m$^{2}$ \cite{Ishikawa99},
not sufficiently large for our purpose. Considering that the lattice
distortion by the small $A$-site ions helps the domain size larger we newly
synthesized the Eu$_{1/2}$Ca$_{3/2}$MnO$_{4}$ (ECMO) compound where the
average size of $A$-site ions ($\sim 1.165$ \AA ) is much smaller than that
in LSMO ($\sim 1.285$ \AA ) \cite{tomioka04}. The $T_{CO-OO}$ for the ECMO\
compound is as high as $\sim $ 325 K perhaps due to the narrower bandwidth,
which is identified from anomalies in resistivity and magnetization, and a
superlattice peak in the electron diffraction pattern (EDP). The $T_{N}$ is
estimated to be 120 - 130 K far below $T_{CO-OO}$ \cite{note1}. Using the
polarized microscope we could identify the stripe-type domain pattern in the
cleaved $ab$ surface. While the typical widths of domains are 10 - 50 $\mu $%
m, we succeeded in finding an exceptionally large single domain with $\sim $
300 $\mu $m width, which enabled us to perform the polarization-dependent
optical measurement for the CO-OO phase.

A Eu$_{1/2}$Ca$_{3/2}$MnO$_{4}$ single crystal with half-doping was grown by
the floating zone method. For the reflectivity measurement of a single
domain, we used the microscope-equipped spectrometers with the beam size, $%
\sim $150 $\times $ 100 $\mu $m$^{2}$. With proper polarizers the probed
light is polarized along the optical axis ($a^{\prime }$ or $b^{\prime }$),
which is rotated by 45 degree from the crystal axes of the tetragonal cell.
The EDP experiment on this material has revealed that the $a^{\prime }$ ($%
b^{\prime }$) axis chosen for our measurement should correspond to the chain
(stripe) direction. $T$-dependent reflectivity spectra $R(\omega )$ at
nearly normal incidence were measured in a photon energy region from 0.06 to
5 eV with the temperature variation from 10 to 430 K. Since the anisotropic
feature in $R(\omega )$ is terminated in the charge transfer excitation
region above 2.5 eV, $T$-dependent $R(\omega )$ below 5 eV are connected
smoothly to room temperature $R(\omega )$ above 5 to 40 eV obtained with a
twinned sample. Optical conductivity spectra $\sigma (\omega )$ were
obtained from the measured $R(\omega )$ using the Kramers-Kronig
transformation.

\begin{figure}[tbp]
\includegraphics[width=0.4\textwidth]{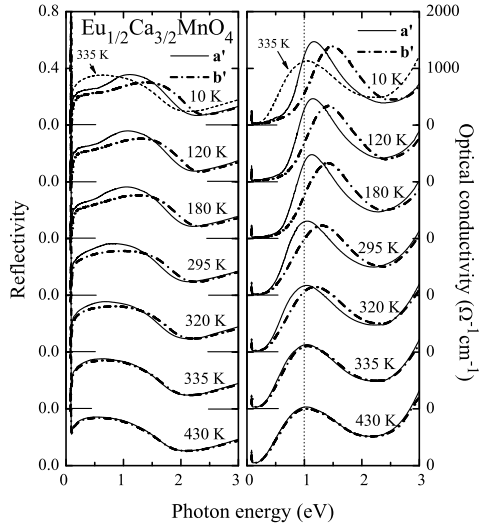}
\caption{Reflectivity spectra (left panels) and optical conductivity spectra
(right panels) of Eu$_{1/2}$Ca$_{3/2}$MnO$_{4}$ along the $a^{\prime }$ and $%
b^{\prime }$ axes. The panels at each temperature are shifted up in
sequence. The 335 K data are included in the top panels to clarify the
temperature-dependence. The dotted line in the right panels is for guidance.}
\end{figure}

Figure 2 shows the reflectivity and the optical conductivity spectra with
polarization dependence at various $T$. Both of optical spectra along $%
a^{\prime }$ and $b^{\prime }$ axes show a similar trend with $T$,
exhibiting significant changes related to the CO-OO. As $T$ is lowered below 
$T_{CO-OO}$, the reflectivity level is suppressed in a low energy region
below $\sim $ 1 eV, while enhanced in the higher energy region. The $\sigma
(\omega )$ calculated from the measured $R(\omega )$ exhibits that a
distinct absorption below 2 eV, which is associated with the electron
hopping from Mn$^{3+}$ (occupied $e_{g}$) to Mn$^{4+}$ (unoccupied $e_{g}$) 
\cite{okimoto98}, shifts to a higher energy with the sharper spectral shape
at lower $T$ below $T_{CO-OO}$, leading to a larger optical gap. The overall
spectral feature is in accord with the previous results for twinned LSMO\
samples \cite{Ishikawa99,jung00}, and the related theoretical calculations 
\cite{CuocoNoceOles02,BalaHorsch05}. Apart from this common $T$-dependence,
we could discern a clear development of anisotropy in optical spectra below $%
T_{CO-OO}$. The reflectivity spectra at 335 K (above $T_{CO-OO}$) appear to
be nearly identical for both polarizations. In contrast, at 320 K (below $%
T_{CO-OO}$) the distinct anisotropy is identified; the level of reflectivity
in the $b^{\prime }$ axis is lower below 1.3 eV, while higher above 1.3 eV,
than in the $a^{\prime }$-axis. The difference between $R(\omega )$ along
the two axes is larger at lower $T$. Similarly, the $\sigma (\omega )$
spectra polarized along the $b^{\prime }$-axis are distinguished from those
along the $a^{\prime }$-axis below $T_{CO-OO}$, exhibiting the higher energy
of absorption band and the larger optical gap.

\begin{figure}[tbp]
\includegraphics[width=0.35\textwidth]{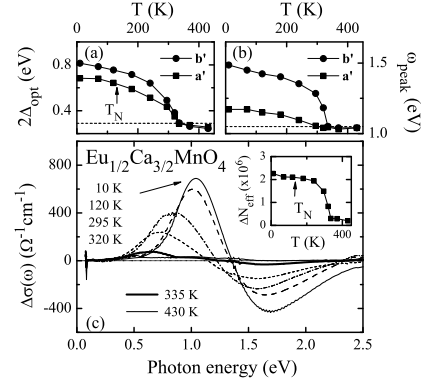}
\caption{(a) Optical gap, 2$\Delta _{\mathrm{opt}}$ and (b) absorption peak
below 2 eV, $\omega _{\mathrm{peak}}$ along the $a^{\prime }$ and $%
b^{\prime }$ axes. (c) Differential conductivity $\Delta \sigma (%
\omega )=$ $\sigma _{a^{\prime }}(\omega )-%
\sigma _{b^{\prime }}(\omega )$ at various temperatures. Inset
displays the temperature dependence of $\Delta N_{\mathrm{eff}}(\omega %
=1.3$ eV$)$. The unit is $\Omega ^{-1}$ cm$^{-2}$.}
\end{figure}

To clarify the in-plane anisotropy, we display the polarization dependence
of the optical gap, 2$\Delta _{\text{opt}}$, and the absorption position
below 2 eV, $\omega _{\text{peak}}$, in Figs. 3(a) and 3(b). The 2$\Delta _{%
\text{opt}}$ values were estimated from crossing points of abscissa with
linear extrapolations of $\sigma (\omega )$. At $T<T_{CO-OO}$ both the
quantities have larger values along the $b^{\prime }$-axis than the $%
a^{\prime }$-axis. The anisotropic difference in 2$\Delta _{\text{opt}}$ and 
$\omega _{\text{peak}}$ are estimated to be $\sim $ 0.12 eV and $\sim $ 0.35
eV, respectively, at the lowest temperature, and hence the degree of
anisotropy is as large as 20 - 30 \%. It is noted that such sizable
anisotropy cannot originate from the orthorhombic crystal structure alone.
While the phonon spectra reveal the lowering of crystal symmetry below $%
T_{CO-OO}$ \cite{Ishikawa99,Yamamoto00}, the degree of orthorhombicity is
too small ($b^{\prime }/a^{\prime }-1<0.01$) to give rise to the significant
electronic anisotropy observed here. In fact, the observed anisotropy above $%
T_{CO-OO}$, which is supposed to come from the inherent lattice
orthorhombicity apart from CO-OO, is extremely small.

Notably, the electronic anisotropy unveiled by the experiments is in good
agreement with the theoretical prediction as shown in Fig. 1. Because the $%
a^{\prime }$ ($b^{\prime }$) axis corresponds to the chain (stripe)
direction, our combined experimental and theoretical studies unambiguously
reveal the formation of the smaller energy gap state along the chain
direction. This finding could be a strong evidence of more effective
electron hopping along the chain direction in the CO-OO phase. Clearly the
in-plane anisotropy identified below $T_{CO-OO}$ is attributed to the
unidirectional zigzag chain arrangement of the $e_{g}$ orbitals accompanied
by the charge ordering. The orbital ordering and the associated anisotropic
transfer integrals leading to the interference of various hopping processes
play a crucial role in the electronic structure of the CO-OO phase. The
direction of orbital stripe (or orbital chain) is closely related to that of
the optical anisotropy. The different optical gap value revealed in our
spectra is expected to lead to the high anisotropy in dc conductivity.
Applying an external electric field might change the direction of the
orbital stripe and the corresponding direction of optical anisotropy. In
this sense the orbital stripe might work as a parameter of a sort of optical
polarization-related device.

Finally we discuss the correlation between the anisotropy and the magnetic
correlation by tracing their $T$-dependent evolutions. To get some insight
into the anisotropic feature in $\sigma (\omega )$ below 2.5 eV, we
estimated the effective spectral weight by integration of $\sigma (\omega )$%
, $N_{\text{eff}}(\omega _{c})=$ $\tint\nolimits_{0}^{\omega _{c}}d\omega
^{\prime }\sigma (\omega ^{\prime })$, and found that $N_{\text{eff}}(\omega
_{c}=2.5$ eV$)$ are nearly identical in both axes. This indicates that the
optical anisotropy observed in the CO-OO state is associated with the
distribution of the local spectral weight below 2.5 eV. To clarify this, we
display the differential conductivity spectra $\Delta \sigma (\omega )=$ $%
\sigma _{a^{\prime }}(\omega )-\sigma _{b^{\prime }}(\omega )$ at various $T$
in Fig. 3(c). While the $\Delta \sigma (\omega )$ above $T_{CO-OO}$ exhibits
quite a small spectral weight with negligible $T$-dependence, the distinct
two peaks with opposite signs, which originate from the accumulation of
lower-energy spectral weight along the $a^{\prime }$ axis, develop
significantly below $T_{CO-OO}$. We also estimated the differential
effective spectral weight $\Delta N_{\text{eff}}$= $\tint\nolimits_{0}^{%
\omega _{c}}d\omega ^{\prime }\Delta \sigma (\omega ^{\prime })$ with $%
\omega _{c}=1.3$ eV, where the $\sigma (\omega )$ in both axes coincide at
the lowest $T$. As shown in the inset of Fig. 4(c), $\Delta N_{\text{eff}}$
exhibits a sizable value with a steep increase below $T_{CO-OO}$, and then
appears to be rather saturated at lower $T$, not exhibiting a distinct
change near $T_{N}$. From this it appears that the three-dimensional (3D)
long-range spin ordering does not affect the optical anisotropy
significantly. When the orbital ordering occurs, the short-range magnetic
correlation could be presumably formed along the zigzag chain even above $%
T_{N}$, which have been identified by the neutron diffraction measurement
for some CO-OO manganites \cite{sternlieb96,Larochelle05,ye05}. No
significant change of the anisotropy near $T_{N}$ implies that this
ferromagnetic fluctuation along the chain could be effective for the
formation of the quasi-1D electron hopping, irrespective of the onset of 3D
long-range magnetic order. This may be indicative of the predominant role of
the orbital ordering in the magnetic correlation.

In summary, we have presented the experimental and theoretical studies on
the optical spectra in the charge/orbital ordered plane with polarization
dependence. The in-plane anisotropy is attributed to the peculiar orbital
arrangement accompanied by the magnetic correlation.

We thank D. I. Khomskii, Y. Okimoto, M. Uchida, R. Mathieu, X. Z. Yu, and Y.
Motome for useful discussions.


\begin{thebibliography}{99}
\bibitem{note1} According to the recent x-ray scattering experiment, the $%
T_{N}$ for the Pr$_{1/2}$Ca$_{3/2}$MnO$_{4}$\ compound with the nearly same $%
T_{CO-OO}$ is found to be 125 K.
\end{thebibliography}
\end{document}